\documentclass[runningheads]{llncs}
\usepackage[T1]{fontenc}
\usepackage{multirow}
\usepackage{booktabs}
\usepackage{siunitx}
\usepackage{comment}
\usepackage[ruled,vlined]{algorithm2e}
\SetCommentSty{textnormal}
\usepackage{graphicx}
\begin{document}
\title{Hilbert Forest in the SISAP 2025 Indexing Challenge}
\author{
Yasunobu Imamura\inst{1}\orcidID{0009-0005-7472-5754} \and
Takeshi Shinohara\inst{2}\orcidID{0000-0002-7451-7374} \and
Naoya Higuchi\inst{3}\orcidID{0009-0004-2377-1681} \and 
Kouichi Hirata\inst{2}\orcidID{0000-0003-0814-8395} \and
Tetsuji Kuboyama\inst{4}\orcidID{0000-0003-1590-0231}}
\authorrunning{Y. Imamura et al.}
\institute{THIRD INC., Shinjuku, Tokyo, Japan \and
Kyushu Institute of Technology, Kawazu 680-4, Iizuka, Japan \\
\email{shino.kyutech@gmail.com, hirata@ai.kyutech.ac.jp} \and
Sojo University, Ikeda 4-22-1, Nishi-ku, Kumamoto, Japan\\
\email{nac24nh@gmail.com} \and
Gakushuin University, Mejiro 1-5-1, Toshima, Tokyo, Japan \\
\email{ori-sisap2025@tk.cc.gakushuin.ac.jp}
}
\maketitle              % typeset the header of the contribution

\begin{abstract}

We report our participation in the SISAP 2025 Indexing Challenge using a novel indexing technique called the Hilbert forest. The method is based on the fast Hilbert sort algorithm, which efficiently orders high-dimensional points along a Hilbert space-filling curve, and constructs multiple Hilbert trees to support approximate nearest neighbor search. We submitted implementations to both Task 1 (approximate search on the PUBMED23 dataset) and Task 2 (k-nearest neighbor graph construction on the GOOAQ dataset) under the official resource constraints of 16 GB RAM and 8 CPU cores. The Hilbert forest demonstrated competitive performance in Task 1 and achieved the fastest construction time in Task 2 while satisfying the required recall levels. These results highlight the practical effectiveness of Hilbert order–based indexing under strict memory limitations. 

\end{abstract}

\section{Introduction}

The SISAP Indexing Challenge 2025~\cite{overview2025sisap} provided a common platform for evaluating indexing techniques under strict computational resource constraints. 
This paper reports our participation in the challenge using the \emph{Hilbert forest}, a recently developed indexing method for approximate $k$-nearest neighbor (ANN) search. 
The Hilbert forest is based on the fast Hilbert sort, which orders high-dimensional points along a Hilbert space-filling curve and builds multiple Hilbert trees to support efficient approximate search.

The source code of the Hilbert forest, as used in the challenge, is publicly available~\cite{hf_github}, and the official challenge results are accessible on GitHub~\cite{SISAP2025_results}. 
We briefly describe the core ideas behind the Hilbert forest, summarize its application to the challenge tasks, and report the achieved performance. 
While a more detailed technical analysis will be presented in future work, this paper serves as a record of our participation and the outcomes of the challenge.

\section{Hilbert Sort and the Hilbert Forest}

\emph{Hilbert sort} refers to reordering a set of points in high-dimensional space in the order along a Hilbert space-filling curve (hereafter, Hilbert order). Using a previously developed algorithm~\cite{hilbert_sort}, we can efficiently sort $n$ points in average $O(n \log n)$ time, independent of dimensionality. Since the Hilbert order defines a total ordering, binary search applies to the sorted sequence. Utilizing the idea of the fast Hilbert sort algorithm, we can locate the point in a set of $n$ points that is closest to a given point $p$ in Hilbert order in average $O(\log n)$ time.

A \emph{Hilbert tree} is a binary search tree built with a point sequence ordered by Hilbert sort. It can be constructed in average $O(n \log n)$ time.

Points that are close in Hilbert order tend to be close in the original space as well, although the converse is not always true. However, due to the super-exponential increase in the number of possible Hilbert curves in higher dimensions, any pair of spatially close points is likely to be adjacent in at least one such Hilbert order.

A \emph{Hilbert forest} is an index structure composed of multiple Hilbert trees, enabling approximate nearest neighbor search. Using more trees improves recall but increases both memory usage and computation time.

\section{Task 1: Approximate Neighbor Search on PUBMED23}

Task 1 deals with the PUBMED23 dataset, consisting of 23 million vectors in 384 dimensions. The goal is to minimize the query throughput time for 10,000 queries under the constraint of RAM 16GB and 8 CPU cores, while keeping recall@30 > 0.7.

In this section, we describe the index structure, search method, and preliminary experimental results obtained in the development environment.

\subsection{Index Structure and Search Method}

The index employs a two-stage filtering approach: coarse candidate selection using a Hilbert Forest, followed by further fine filtering with sketches. Final candidate selection is performed based on vector distance.

A full binary search tree constructed on the entire dataset would exceed 400MB per Hilbert tree. To stay within the 16GB RAM limit and leave room for sketches and quantized vectors used in later stages, we compressed each tree to about 76MB by replacing subtrees containing about 100 points with leaf nodes. The number of trees in the forest is at most 160.

To accelerate the filtering step, we avoid processing each query independently. Instead, we simulate the Hilbert sort for all queries collectively and compute their positions in Hilbert order. One Hilbert order is selected as the master order, and all vectors and sketches are rearranged in memory accordingly. This improves memory access locality in the second filtering stage. Additionally, we optionally include several neighbors around the selected candidates in the master order to improve recall.

For the second-stage filtering, we used 384-bit sketches, striking a balance between discriminative power and memory usage. The total size of the sketches is approximately 1.1 GB (23M × 384 bits). Hamming distance was used as the selection criterion.

The original vectors are represented by 32-bit float values, which amount to approximately 36 GB, and cannot be fully loaded into RAM. Therefore, we quantized the vectors to 4-bit integers. To reduce memory usage further, one bit is shared with the sketch, compressing the combined memory footprint of sketches and quantized vectors to about 4.5 GB. For accuracy, queries are not quantized during the final distance computation.

\subsubsection{Key Hyperparameters}

\begin{itemize}
  \item Number of trees in the Hilbert Forest: $n$
  \item Number of candidates extracted per query from each tree: $k_1$
  \item Number of candidates selected per query by sketches: $k_2$
  \item Number of additional neighbors from the master order per candidate: $h$
\end{itemize}

The total number of candidates selected by the forest (including duplicates) for each query is $n \times k_1$. The number of candidates examined in the second-stage filtering is approximately $k_2 \times (2h + 1)$ per query.

%\subsection{Search Flow} 
We summarize the entire procedure of our approximate $k$-NN search method using the Hilbert forest in Algorithm~\ref{alg:task1}.

\vspace{1em}

\begin{algorithm}[h]
\caption{Approximate $k$-NN Search Using Hilbert Forest}
\label{alg:task1}
\KwIn{Queries $q[0], \ldots, q[Q - 1]$}
\KwOut{Result sets $S[0], \ldots, q[Q - 1]$}

% === Preprocessing ===
\tcp{Preprocessing steps}
Perform vector quantization of the dataset\;
Generate binary sketches for all vectors\;
Create a master order using Hilbert sorting and adjust memory layout for vectors and sketches\;
Build $n$ Hilbert trees with randomized axis orders\;

% === Initialization ===
\tcp{Initialize candidate sets for each query}
\For{$i \leftarrow 0$ \KwTo $Q - 1$}{
    $C_1[i] \leftarrow \emptyset$\;
}

% === First-stage filtering ===
\tcp{Use Hilbert forest to collect candidates}
\For{$t \leftarrow 0$ \KwTo $ntrees - 1$}{
    simulate Hilbert sort of $q[0]$ to $q[Q - 1]$ using $tree[t]$\;
    \For{$i \leftarrow 0$ \KwTo $Q - 1$}{
        extract $k_1$ candidates near $q[i]$'s position and add to $C_1[i]$\;
    }
}

% === Second-stage filtering ===
\tcp{Filter candidates using Hamming distance of sketches}
\For{$i \leftarrow 0$ \KwTo $Q - 1$}{
    $C_2[i] \leftarrow$ select top-$k_2$ candidates from $C_1[i]$ using Hamming distance\;
}

% === Final selection ===
\tcp{Optionally expand and select final results using full vector distance}
\For{$i \leftarrow 0$ \KwTo $Q - 1$}{
    expand $C_2[i]$ using the master order (e.g., include $h$ neighbors on each side of each candidate)\;
    $S[i] \leftarrow$ select top-$k$ candidates from $C_2[i]$ using vector distance\;
}
\end{algorithm}

\subsection{Experimental Results}

The experiments were conducted on a development machine with a Ryzen 9 3950X CPU and 128GB RAM. The actual runs were performed in a container limited to 8 CPU cores and 16GB RAM using WSL2 on Windows 11.

Index construction times were as follows:
\begin{itemize}
  \item Hilbert Forest (160 trees): 38m56s
  \item Sketch generation: 5s
  \item Vector quantization: 2m32s
  \item Master sorting: 14s
\end{itemize}
Total preprocessing time: 2594 seconds (approximately 43 minutes)

Results for 16 hyperparameter combinations are summarized in Table~\ref{tab:task1}.  

\begin{table}[h]
  \centering
  \caption{Recall and Search Time for Task1}
  \begin{tabular}{r@{~~} r@{~~} r@{~~} r r r}
    \toprule
$n${~} & $k_1${~} & $k_2${~} & $h$ & recall (\%) & time (sec) \\
\midrule
160 & 1420 & 370 & 2 & 72.9{~~~} & 54.9{~~~} \\
160 & 1420 & 360 & 2 & 72.8{~~~} & 18.0{~~~} \\
160 & 1300 & 350 & 2 & 71.9{~~~} & 17.9{~~~} \\
160 & 1300 & 340 & 2 & 71.8{~~~} & 16.1{~~~} \\
160 & 1200 & 330 & 2 & 71.0{~~~} & 14.9{~~~} \\
160 & 1200 & 320 & 2 & 70.9{~~~} & 17.3{~~~} \\
160 & 1100 & 310 & 2 & 70.0{~~~} & 13.6{~~~} \\
160 & 1100 & 300 & 2 & 69.9{~~~} & 15.7{~~~} \\
\midrule
120 & 4000 & 1000 & 2 & 79.1{~~~} & 29.4{~~~} \\
120 & 3200 & 1000 & 2 & 77.2{~~~} & 25.8{~~~} \\
120 & 2800 & 1000 & 2 & 76.0{~~~} & 23.6{~~~} \\
120 & 2400 & 1000 & 2 & 74.5{~~~} & 23.8{~~~} \\
120 & 2000 & 1000 & 2 & 72.6{~~~} & 19.4{~~~} \\
120 & 1800 & 1000 & 2 & 71.5{~~~} & 18.3{~~~} \\
120 & 1600 & 1000 & 2 & 70.3{~~~} & 19.6{~~~} \\
120 & 1600 & 800 & 2 & 70.0{~~~} & 15.6{~~~} \\
\bottomrule
\end{tabular}
\label{tab:task1}
\end{table}

\section{Task 2: k-Nearest Neighbor Graph Construction on GOOAQ}

Task 2 targets the GOOAQ dataset, which consists of 3 million vectors of 384 dimensions. The goal is to construct a $k$-nearest neighbor ($k$-NN) graph for all data points under the constraint of 16GB RAM and 8 CPU cores, while achieving recall@15 > 0.8. The primary evaluation criterion is the total construction time.

\subsection{Hilbert Forest-Based Graph Construction}

Since every data point is also a query in this task, a Hilbert-ordered sequence can be used to select neighboring candidates based on proximity in the sorted order. Therefore, unlike in Task 1, the binary search tree component of each Hilbert tree is not needed. Once candidates are extracted using a Hilbert order, the sorted sequence can be discarded. Even when using multiple Hilbert orders, memory consumption remains constant, with only the computation time increasing.

\vspace{1cm}

\subsubsection{Key Hyperparameters}

\begin{itemize}
  \item Number of Hilbert sorts used: $n$
  \item Number of candidates extracted per point from each sort: $k_1$
  \item Number of candidates selected per point by sketches: $k_2$
\end{itemize}

%\subsection{Construction Flow}

We give the outline of the graph construction in Algorithm~\ref{alg:task2}. 

\begin{algorithm}[h]
\caption{Approximate $k$-NN Graph Construction for Task 2}
\label{alg:task2}

% === Preprocessing ===
\tcp{Preprocessing steps}
Vector quantization\;
Generate binary sketches for all vectors\;

\textbf{Input:} Dataset of size $N$\\
\textbf{Output:} Approximate $k$-NN graph $S$.

% Initialization
\tcp{Initialize candidate sets}
\For{$i \gets 0$ \textbf{to} $N-1$}{
  $C_1[i] \gets \emptyset$ 
}

\tcp{First-stage filtering (using Hilbert sorts)}
\For{$t \gets 0$ \textbf{to} $n-1$}{
  randomly permute coordinate axes\;
  compute Hilbert order of all points\;
  \For{$i \gets 0$ \textbf{to} $N-1$}{
    $j \gets$ index of the $i$-th point in the Hilbert order\;
    extract $k_1$ neighbors around position $i$ in the order\;
    append their indices to $C_1[j]$\;
  }
}

\tcp{Second-stage filtering (using sketches)}
\For{$i \gets 0$ \textbf{to} $N-1$}{
  $C_2[i] \gets$ select top-$k_2$ candidates from $C_1[i]$ using Hamming distance\;
}

\tcp{Final selection (using vector distances)}
\For{$i \gets 0$ \textbf{to} $N-1$}{
  $S[i] \gets$ select top-15 from $C_2[i]$ using vector distance\;
}

\end{algorithm}

\subsection{Experimental Results}

The Table~\ref{tab:task2} below summarizes performance under different numbers of Hilbert sorts:

\begin{table}[htbp]
  \centering
  \caption{Construction Performance for Task2}
\begin{tabular}{r@{~~} r@{~~} r@{~~} r@{~~} r}
    \toprule
Time (sec) & Recall (\%) & $n${~} & $k_1$ & $k_2$ \\
\midrule
74{~~~} & 80.5{~~~} & 80 & 96 & 60 \\
109{~~~} & 85.5{~~~} & 112 & 106 & 75 \\
164{~~~} & 90.5{~~~} & 160 & 130 & 100 \\
330{~~~} & 95.5{~~~} & 280 & 168 & 150 \\
856{~~~} & 98.5{~~~} & 720 & 170 & 300 \\
\bottomrule
\end{tabular}
\label{tab:task2}
\end{table}

A minimum recall of 80\% can be achieved in 74 seconds. Moreover, a graph with 95\% recall can be constructed in just 5 minutes and 30 seconds.

\section{Conclusion}

For Task 1, due to memory limitations, it was difficult to increase the number of trees in the Hilbert forest, and thus achieving very fast search performance was not possible. However, it has been confirmed that under conditions with sufficient memory, significantly faster searches can be realized. In contrast, for Task 2, since it is not necessary to keep all Hilbert forest trees resident in memory, and a single Hilbert order sequence can be used to select candidates from the entire dataset, very fast processing was achieved.

The approximate nearest neighbor search method using the proposed Hilbert forest, although still in the early stages of development, shows promise in providing an effective indexing structure and search strategy for high-dimensional data. In particular, the combination of multiple Hilbert orders leveraging the properties of space-filling curves is meaningful in achieving a balance between recall and speed.

Going forward, we aim to further develop this method into a practical high-dimensional indexing technique by applying it to more diverse datasets and constrained environments, optimizing parameters at each stage, and conducting comparative evaluations with other filtering methods.

\bibliographystyle{splncs04}
\bibliography{bib}

\end{document}